\documentclass[english,aps,twocolumn]{revtex4}
\usepackage[T1]{fontenc}
\usepackage[latin9]{inputenc}
\usepackage{amsmath}
\usepackage{amssymb}
\usepackage{babel}

\begin{document}

%\preprint{This line only printed with preprint option}

\title{The sign of the overlap of HFB wave functions}

\author{L.M. Robledo}

\email{luis.robledo@uam.es}

\affiliation{Dep. Física Teórica C-XI, Facultad de Ciencias, Universidad Autónoma
de Madrid, 28049 Madrid, Spain}

\begin{abstract}
The problem of how to compute accurately and efficiently the sign
of the overlap between two general HFB wave functions is addressed.
The results obtained can easily be extrapolated to the evaluation
of the sign of the trace of a density operator exponential of one
body operators. 
\end{abstract}
\maketitle

\section{introduction}

Beyond mean field calculations are becoming very popular \cite{Bender.03}
as they allow a fairly good description of many nuclear state properties
of both the ground state and several kinds of excited states all over
the Nuclide chart. In these calculations, overlaps of Hartree- Fock-
Bogoliubov (HFB) wave functions have to be computed. Standard formulas
\cite{RS.80} involve the square root of a determinant leaving the
sign of the overlap undefined. However, when the HFB states preserve
some kind of discrete symmetry like time reversal or simplex, the
block structure of the matrices involved fixes the sign. This has
been discussed, for instance, in some recent applications of angular
momentum projection (AMP) using axially symmetric and time reversal
preserving intrinsic wave functions \cite{rodriguez.02,valor.00}.
To move forward, HFB wave functions that do not have any spatial symmetry
(triaxial) and also breaking time reversal symmetry have to be considered
in order to incorporate $K\ne0$ configurations. This is the case
to describe, for instance, the ground state of odd-A nuclei. For the
usual time reversal breaking (TRB) mean field wave functions, the
simplex symmetry endows the HFB amplitudes U and V with a common bipartite
structure and the usual arguments used to extract out the sign of
the overlap apply. However, when full triaxial angular momentum projection
of HFB intrinsic states \cite{Bender.08} is considered, the simplex
symmetry is no longer preserved in the evaluation of rotated overlaps
and the determination of the sign becomes more difficult. A general
solution to the sign problem was given in Ref. \cite{neergard.83},
where it was shown that the overlap, including the sign, can be computed
from the pairwise degenerate eigenvalues of a non-hermitian matrix.
Handling the eigenvalues of non-hermitian matrices is a difficult
task \cite{Golub.96}, that increases its complexity if the pairwise
degenerate eigenvalues have to be obtained numerically without any
symmetry enforcing degeneracy, as is the case with HFB wave functions
breaking simplex. Neergard's method has been used along with small
configuration spaces \cite{Schmid.04} but in the majority of the
calculations continuity arguments are used (see Refs. \cite{Hara.82,Oi.05,Bender.08}
for recent examples) in spite of the difficulties with that procedure.
The same sign problem is also present in the evaluation of the trace
of statistical density operators \cite{Lang.93,Rossignoli.94}. In
this case, however, Neergard's method has not been implemented up
to date, leaving as the only choice the continuity method in such
finite temperature calculations. The same difficulty also applies
to the recently proposed method to compute multiquasiparticle overlaps
that relays on the statistical Wick's theorem \cite{SPM.07}. Recently,
\cite{Taka.08} the group structure of the unitary Bogoliubov transformation
has been discussed, as well as its implications in the relative phase
between two HFB wave functions. However, its practical implications are
still unclear.

In this paper, I will introduce a new way to compute the overlap of two
HFB wave functions based on the concept of fermion coherent states
\cite{Berezin.66,Blaizot.85,Negele.88}. The new formula involves
a quantity similar to the determinant called pfaffian of a skew-symmetric
matrix. The advantage of the proposed method is that the numerical
evaluation of the pfaffian is simple and lacks the problems previously
mentioned about pairwise degenerate eigenvalues. Another advantage
of the present formulation is its applicability to the evaluation
of the trace of density operators like the ones found in applications
of the auxiliary-field shell model Monte Carlo\cite{Lang.93} or
symmetry restoration at finite temperature\cite{Rossignoli.94}.
A reliable determination of the sign of the norm can also be useful
in to order to pin down the location of the zeros of the HFB overlaps
\cite{Oi.05}. This determination would eventually be useful to get
rid of the so called {}``pole problem'' that plagues present beyond
mean field calculations.

\section{Overlaps and traces}

\subsection{Preliminaries}

Let $|\phi_{0}\rangle$ and $|\phi_{1}\rangle$ be two HFB wave functions
defined in terms of a set of single particle creation and annihilation
operators $a_{k}^{+}$ and $a_{k}$ that are assumed to be related
by hermitian conjugation and also to satisfy fermion commutation relations.
The HFB wave functions, in the Thouless representation \cite{RS.80,Blaizot.85},
 are given by \begin{equation}
|\phi_{i}\rangle=\exp\left(\frac{1}{2}
\sum_{kk'}M_{kk'}^{(i)}a_{k}^{+}a_{k'}^{+}\right)|0\rangle\label{eq:HFBWF}\end{equation}
where the skew-symmetric matrices\[
M^{(i)}=(V_{i}U_{i}^{-1})^{*}\]
are defined in terms of the $U_{i}$ and $V_{i}$ coefficients of
the Bogoliubov transformations defining the HFB wave functions and
$|0\rangle$ is the true vacuum. The arbitrary phase that can always
be associated with a vector state in quantum mechanics has been implicitly
fixed in the definition of Eq. (\ref{eq:HFBWF}) by requiring 
$\langle0|\phi_{i}\rangle=1$. Ways to enforce this normalization 
for general HFB wave functions are discussed, for instance in 
Refs. \cite{RS.80,Blaizot.85}. In the event of having $\langle0|\phi_{i}\rangle=0$
(as a consequence of divergent $M^{(i)}$ and/or zero occupancies) the
best practical strategy is to use another reference wave function 
instead of the true vacuum $|0\rangle$. The new reference HFB wave function
$|\bar{\phi}\rangle$ has to be conveniently chosen as to stay close to both $|\phi_{i}\rangle$
(for instance by taking a wave function with similar deformation parameters 
as those of $|\phi_{i}\rangle$). The matrices $\bar{M}^{(i)}$ referred to $|\bar{\phi}\rangle$
can be straightforwardly computed in terms of the previous quantities and the 
Bogoliubov transformation amplitudes of the reference state. In the rare event
of not finding a convenient reference wave function $|\bar{\phi}\rangle$
a regularization procedure to handle the divergent $\bar{M}^{(i)}$ 
matrix elements (or the zero occupancies) is in order. In this case,
the expressions get more involved and a detailed account is deferred to
a forthcoming publication. Another way to deal with that problem is
presented in Ref. \cite{neergard.83} but the resulting expressions are
rather involved.

Let me now introduce fermion coherent states $|\mathbf{z}\rangle,$
which are parametrized in terms of the anticommuting elements
$z_{k}$ and $z_{k}^{*}$ of a
Grassmann algebra \cite{Berezin.66,Blaizot.85,Negele.88,Ohnuki.78}
and fulfilling the equations 
\begin{equation}
a_{k}|\mathbf{z}\rangle=z_{k}|\mathbf{z}\rangle\label{eq:z}\end{equation}
and
\begin{equation}
\langle\mathbf{z}|a_{k}^{+}=z_{k}^{*}\langle\mathbf{z}|\label{eq:zstar}\end{equation}
From the above definition is clear that $|\mathbf{z}\rangle$ is a
right eigenstate of the annihilation operator $a_{k}$ with eigenvalue
$z_{k}$ whereas $\langle\mathbf{z}|$ is a left eigenvector of $a_{k}^{+}$
with eigenvalue $z_{k}^{*}$(The notation used for the members of
the Grassmann algebra is the usual one but can be a little misleading
as $z_{k}^{*}$ is not connected to $z_{k}$ by complex conjugation).
The coherent states satisfy a closure relation 
\begin{equation}
\openone=\int d\mu(\mathbf{z})|\mathbf{z}\rangle\langle\mathbf{z}|\label{eq:closure}\end{equation}
where the metric of the integral is given by $d\mu(\mathbf{z})=e^{-\mathbf{z}^{*}\mathbf{z}}\prod_{k}dz_{k}^{*}dz_{k}$.
These and other relevant definitions and properties of fermion coherent
states can be found in many textbook or in the original literature
\cite{Berezin.66,Blaizot.85,Negele.88,Ohnuki.78}. 

\subsection{Evaluation of the overlap}

To compute the overlap $\langle\phi_{0}|\phi_{1}\rangle$, the closure
relation of Eq. (\ref{eq:closure}) is inserted to obtain\begin{eqnarray*}
\langle\phi_{0}|\phi_{1}\rangle & = & \int d\mu(\mathbf{z})\langle0|e^{\frac{1}{2}\sum_{kk'}M_{kk'}^{(0)\,*}a_{k'}a_{k}}|\mathbf{z}\rangle\\
 & \times & \langle\mathbf{z}|e^{\frac{1}{2}\sum_{kk'}M_{kk'}^{(1)}a_{k}^{+}a_{k'}^{+}}|0\rangle\end{eqnarray*}
Using now Eqs. (\ref{eq:z}) and (\ref{eq:zstar}) one arrives to
\begin{equation}
\langle\phi_{0}|\phi_{1}\rangle=\int d\mu(\mathbf{z})e^{\frac{1}{2}\sum_{kk'}M_{kk'}^{(0)\,*}z_{k'}z_{k}}e^{\frac{1}{2}\sum_{kk'}M_{kk'}^{(1)}z_{k}^{*}z_{k'}^{*}}\label{eq:Integral}\end{equation}
where the property $|\langle0|\mathbf{z}\rangle|^{2}=1$ is used.
The integral is of the Gaussian type but for Grassmann variables.
The techniques to evaluate this kind of integrals can be found in
many textbooks \cite{Berezin.66,Blaizot.85,Negele.88} but its evaluation
will be carried out explicitly here. The reason is that in order to
determine the sign of the norm we have to be careful with some intermediate
steps. The above integral can be written in a more compact way by
introducing the bipartite skew-symmetric matrix\[
\mathbb{M}_{\mu'\mu}=\left(\begin{array}{cc}
M_{k'k}^{(1)} & -\openone_{k'k}\\
\openone_{k'k} & -M_{k'k}^{(0)\,*}\end{array}\right)\]
and the vector of Grassmann variables $z_{\mu}=(z_{k'}^{*},z_{k'})$
as
\begin{equation}
\langle\phi_{0}|\phi_{1}\rangle=\int\prod_{k}\left(dz_{k}^{*}dz_{k}\right)e^{\frac{1}{2}\sum_{\mu\mu'}z_{\mu'}\mathbb{M}_{\mu'\mu}z_{\mu}}\label{eq:Over0}\end{equation}
The skew-symmetric matrix $\mathbb{M}$ can always be transformed
\cite{Zumino.62} to canonical form by means of a unitary transformation
$U$\[
\mathbb{M}=U\left(\begin{array}{cccccc}
0 & \cdots & 0 & \beta_{1} & 0 & 0\\
\vdots & \ddots & \vdots & 0 & \ddots & 0\\
0 & \cdots & 0 & 0 & 0 & \beta_{N}\\
-\beta_{1} & 0 & 0 & 0 & \cdots & 0\\
0 & \ddots & 0 & \vdots & \ddots & \vdots\\
0 & 0 & -\beta_{N} & 0 & \cdots & 0\end{array}\right)U^{T}=U\mathbb{M}_{c}U^{T}\]
and the $\beta_{1},\ldots,\beta_{N}$ coefficients of the {}``canonical
form'' of the matrix $\mathbb{M}$ are real and positive. Introducing
now the new Grassmann variables $\eta_{\mu}=\sum_{\mu'}(U^{T})_{\mu\mu'}z_{\mu'}$
the exponent in the integrand of Eq. (\ref{eq:Integral}) becomes
\[
\frac{1}{2}\sum_{\mu\mu'}\eta_{\mu}\mathbb{M}_{c\,\mu\mu'}\eta_{\mu'}=\sum_{k=1}^{N}\beta_{k}\eta_{k}^{*}\eta_{k}\]
which is straightforward to integrate. The Jacobian of the transformation
can be shown to be simply $\det(U^{T})=\det(U)$. The remaining integrals
can be performed easily being the result \[
\int d\eta^{*}d\eta e^{\beta\eta^{*}\eta}=-\beta.\]
 The final expression for the overlap is then \[
\langle\phi_{0}|\phi_{1}\rangle=(-1)^{N}\det(U)\prod_{k=1}^{N}\beta_{k}.\]
This expression can be cast in terms of the pfaffian of a skew-symmetric
matrix. The pfaffian of a skew-symmetric matrix (see, for instance,
\cite{Caianello}) is a number obtained out of the matrix elements
of the skew-symmetric matrix in a way quite similar to the one used
to define the determinant (see Appendix A for details and properties
used below). The connection between the product of $\beta_{i}$'s
and the pfaffian is a consequence of Eq. \eqref{eq:pfRC} and reads
$\prod_{k=1}^{N}\beta_{k}=(-1)^{N(N-1)/2}\textrm{pf}(\mathbb{M}_{c})$
where $\textrm{pf}(\mathbb{M}_{c})$ obviously denotes the pfaffian
of $\mathbb{M}_{c}$. Using the property \eqref{eq:PTRP} $\textrm{pf}(\mathbb{M})=\textrm{pf}(U\mathbb{M}_{c}U^{T})=\det(U)\textrm{pf}(\mathbb{M}_{c})$
the final result is obtained,
\begin{equation}
\langle\phi_{0}|\phi_{1}\rangle=s_{N}\textrm{pf}(\mathbb{M})=s_{N}\textrm{pf}\left(\begin{array}{cc}
M^{(1)} & -\openone\\
\openone & -M^{(0)\,*}\end{array}\right)\label{eq:Over_1}\end{equation}
where $s_{N}=(-1)^{N(N+1)/2}$. To make the connection with the standard
formula for the overlap \cite{RS.80} the relation $\textrm{pf}(A)^{2}=\det A$
is used (and this is here where the sign is lost) to write
\begin{equation}
\langle\phi_{0}|\phi_{1}\rangle=\left(\det\left(\begin{array}{cc}
M^{(1)} & -\openone\\
\openone & -M^{(0)\,*}\end{array}\right)\right)^{1/2}\label{eq:Onishi_0}\end{equation}
This expression reduces, by using standard formulas for the determinant
of a bipartite matrix (see below), to 
\begin{equation}
\langle\phi_{0}|\phi_{1}\rangle=\left(\det(\openone-M^{(0)\,*}M^{(1)})\right)^{1/2}\label{eq:Onishi_1}\end{equation}
which is the usual expression for the norm (Onishi formula). Please
notice that in going from Eq. (\ref{eq:Over_1}) to Eq. (\ref{eq:Onishi_0})
the sign present in the first equation is lost as a consequence of
the writing of the square of the pfaffian as a determinant. Also signs
appearing in the manipulations needed to obtain Eq. (\ref{eq:Onishi_1})
have been neglected. We clearly see that the sign problem appears
in the standard formulas because of the wrong implicit use of the
above relation between the pfaffian and the determinant.

If both HFB wave functions $|\phi_{0}\rangle$ and $|\phi_{1}\rangle$
share a common discrete symmetry like simplex or time reversal, then
the matrices $M^{(i)}$ defining them can acquire a common block structure
\[
M^{(i)}=\left(\begin{array}{cc}
0 & \overline{M}^{(i)}\\
-\overline{M}^{(i)T} & 0\end{array}\right)\]
that can be used to simplify the result of Eq. (\ref{eq:Over_1}).
In this case\[
\mathbb{M}=\left(\begin{array}{cccc}
0 & \overline{M}^{(1)} & -\openone & 0\\
-\overline{M}^{(1)T} & 0 & 0 & -\openone\\
\openone & 0 & 0 & -\overline{M}^{(0)*}\\
0 & \openone & \overline{M}^{(0)+} & 0\end{array}\right)\]
By exchanging blocks 2 and
4 we obtain 
\[
\textrm{pf}(\mathbb{M})=(-1)^{N}\textrm{pf}\left(\begin{array}{cccc}
0 & 0 & -\openone & \overline{M}^{(1)}\\
0 & 0 & \overline{M}^{(0)+} & \openone\\
\openone & -\overline{M}^{(0)*} & 0 & 0\\
-\overline{M}^{(1)T} & -\openone & 0 & 0\end{array}\right)\]
that can be evaluated using Eq. (\ref{eq:pfRC}) to give
\begin{equation}
\langle\phi_{0}|\phi_{1}\rangle=\det\left(\begin{array}{cc}
-\openone & \overline{M}^{(1)}\\
\overline{M}^{(0)+} & \openone\end{array}\right)=
\det(\openone + \overline{M}^{(0)+}\overline{M}^{(1)}).\label{eq:Overlap_simplex}
\end{equation}

\subsection{Evaluation of statistical traces}

Now I turn  to the evaluation of the trace of density operators. In
the statistical HFB theory the statistical density operator$\hat{\mathcal{D}}$is
given by the exponential of a one body operator $\hat{\mathcal{D}}=\exp[\frac{1}{2}\sum_{\mu\nu}\gamma_{\mu}\mathcal{R}_{\mu\nu}\gamma_{\nu}]$
where $\gamma_{\mu}$is a shorthand notation for $(\beta_{1},\ldots,\beta_{N},\beta_{1}^{+},\ldots,\beta_{N}^{+})$
and $\mathcal{R}$ is a skew-symmetric matrix of dimension $2N$ characterizing
the density operator (see Ref \cite{Rossignoli.94} for details).
Another way to characterize the density operator is to define how
it transforms quasiparticle creation and annihilation operators $\hat{\mathcal{D}}^{-1}\gamma_{\mu}\hat{\mathcal{D}}=\sum_{\nu}T_{\mu\nu}\gamma_{\nu}$
where the matrix $T=\exp(\sigma\mathcal{R})$ and $\sigma_{\mu\nu}=\{\gamma_{\mu},\gamma_{\nu}\}$.
Introducing the bipartite structure of $T$

\[
T=\left(\begin{array}{cc}
T_{11} & T_{12}\\
T_{21} & T_{22}\end{array}\right)\]
the Balian and Brezin's decomposition \cite{Balian.69} of $\hat{\mathcal{D}}$
is given by
\begin{equation}
\hat{\mathcal{D}}=e^{\frac{1}{2}\sum_{ij}\beta_{i}^{+}X_{ij}\beta_{j}^{+}}e^{-\frac{1}{2}\textrm{Tr[Y]}}e^{\sum_{ij}\beta_{i}^{+}Y_{ij}\beta_{j}}e^{\frac{1}{2}\sum_{ij}\beta_{i}Z_{ij}\beta_{j}}\label{eq:BBdec}\end{equation}
with $X=T_{12}T_{22}^{-1}$ and $Z=T_{22}^{-1}T_{21}$ skew-symmetric
(as a consequence of the relation $T^{T}\sigma T=\sigma$ that $T$
satisfies) and $\exp(-Y)=T_{22}^{T}$. To evaluate the trace of $\hat{\mathcal{D}}$
using fermion coherent states we have to use the formula \cite{Negele.88}
\begin{equation}
\textrm{Tr}(\hat{\mathcal{D}})=\int d\mu(\mathbf{z})\langle-\mathbf{z}|\hat{\mathcal{D}}|\mathbf{z}\rangle\label{eq:TrD}\end{equation}
where $|\mathbf{z}\rangle$ are again a set of fermion coherent states
but chosen this time as eigenstates of the quasiparticle annihilation
operators $\beta_{i}$, i.e. $\beta_{i}|\mathbf{z}\rangle=z_{i}|\mathbf{z}\rangle$.
Using Eq. (\ref{eq:BBdec}) the evaluation of the overlap between
the fermion coherent states gives \begin{eqnarray*}
\langle-\mathbf{z}|\hat{\mathcal{D}}|\mathbf{z}\rangle & = & e^{-\frac{1}{2}\textrm{Tr[Y]}}e^{\frac{1}{2}\sum_{ij}z_{i}^{*}X_{ij}z_{j}^{*}}e^{\frac{1}{2}\sum_{ij}z_{i}Z_{ij}z_{j}}\\
 & \times & \langle-\mathbf{z}|e^{\sum_{ij}\beta_{i}^{+}Y_{ij}\beta_{j}}|\mathbf{z}\rangle\end{eqnarray*}
To evaluate the remaining overlap the standard result $\exp\left(\sum_{ij}\beta_{i}^{+}Y_{ij}\beta_{j}\right)|\mathbf{z}\rangle=|e^{Y}\mathbf{z}\rangle$
used together with $\langle-\mathbf{z}|\mathbf{z}'\rangle=\exp(-\mathbf{z}^{*}\mathbf{z}')$
(see Refs. \cite{Berezin.66,Blaizot.85,Negele.88} ) gives\begin{eqnarray*}
\langle-\mathbf{z}|\hat{\mathcal{D}}|\mathbf{z}\rangle & = & e^{-\frac{1}{2}\textrm{Tr[Y]}}e^{\frac{1}{2}\sum_{ij}z_{i}^{*}X_{ij}z_{j}^{*}}e^{-\sum_{ij}z_{i}^{*}(e^{Y})_{ij}z_{j}}\\
 & \times & e^{\frac{1}{2}\sum_{ij}z_{i}Z_{ij}z_{j}}\end{eqnarray*}
Combining this result with Eq. \eqref{eq:TrD}, the following integral
is obtained\[
\textrm{Tr}(\hat{\mathcal{D}})=e^{-\frac{1}{2}\textrm{Tr[Y]}}\int\prod_{k}\left(dz_{k}^{*}dz_{k}\right)e^{\frac{1}{2}\sum_{\mu\mu'}z_{\mu'}\mathbb{M}_{\mu'\mu}z_{\mu}}\]
where the same notation as in Eq \eqref{eq:Over0} is used. In this
case \[
\mathbb{M}=\left(\begin{array}{cc}
X & -(e^{Y}+\openone)\\
(e^{Y}+\openone)^{T} & Z\end{array}\right)\]
Applying the same considerations as in the evaluation of the overlap
we finally arrive to\[
\textrm{Tr}(\hat{\mathcal{D}})=s_{N}\exp\left(-\frac{1}{2}\textrm{Tr[Y]}\right)\textrm{pf}(\mathbb{M})\]
where $s_{N}=(-1)^{N(N+1)/2}$. Taking into account the relationship
between $X$, $Z$ and $Y$ and the blocks of the matrix $T$ the
above result can be expressed as\begin{eqnarray*}
\textrm{Tr}(\hat{\mathcal{D}}) & = & s_{N}\left(\det T_{22}\right)^{1/2}\\
 & \times & \textrm{pf }\left(\begin{array}{cc}
T_{12}T_{22}^{-1} & -(\left(T_{22}^{T}\right)^{-1}+\openone)\\
(\left(T_{22}\right)^{-1}+\openone) & T_{22}^{-1}T_{21}\end{array}\right)\end{eqnarray*}
The introduction of $\left(\det T_{22}\right)^{1/2}$ in place of
$\exp\left(-\frac{1}{2}\textrm{Tr[Y]}\right)$ can lead to the (right)
conclusion that a sign indeterminacy has been introduced in the expression
of the trace. The definition of $\hat{\mathcal{D}}$ in terms of the
transformation matrix $T$ leaves a phase open in the definition of
the density operator which is also present in the expression of Eq.
\eqref{eq:BBdec}. A way to fix the phase is to require some condition
like, for instance, the realness and positiveness of $\langle\phi_{0}|\hat{\mathcal{D}}|\phi_{0}\rangle=\left(\det T_{22}\right)^{1/2}$
where $|\phi_{0}\rangle$ is the vacuum of the quasiparticle operators
$\beta_{i}$ entering in the definition of $\hat{\mathcal{D}}$. This
condition implies the replacement of $\left(\det T_{22}\right)^{1/2}$
by its modulus. Using property \eqref{eq:PTRP} of the pfaffian the
final result is obtained\begin{eqnarray}
\textrm{Tr}(\hat{\mathcal{D}}) & = & s_{N}\frac{\left|\det T_{22}\right|^{1/2}}{\det T_{22}}\label{eq:TrDFinal}\\
 & \times & \textrm{pf }\left(\begin{array}{cc}
T_{12}T_{22}^{-1} & -(T_{22}^{T}+\openone)\\
(T_{22}+\openone) & T_{21}T_{22}^{T}\end{array}\right)\end{eqnarray}
This result is apparently quite different from the standard one of
\cite{Lang.93,Rossignoli.94}, but after some tedious manipulations
(see appendix B) one can obtain the usual result.

\section{Conclusions}

I have used the technique of fermion coherent states
to compute unambiguously the sign of the overlap of two HFB wave functions.
The result given in terms of pfaffians is simpler to implement than
previous considerations \cite{neergard.83} based on pairwise degenerate
eigenvalues of a general matrix and it is free from the uncertainties
of other methods based on continuity arguments. Indications on how
to evaluate efficiently the pfaffian are also given. Hopefully, this
new method will help to simplify the implementation of ambitious projects
like triaxial angular momentum projection. On the other hand, the
method used is straightforwardly extended to the evaluation of
the sign of the trace of statistical density operators which is a
new result not considered previously in the literature. 

\begin{acknowledgments}
Work supported in part by MEC (FPA2007-66069) and by the Consolider-Ingenio
2010 program CPAN (CSD2007-00042)
\end{acknowledgments}
\appendix

\section{Definition, basic properties and numerical evaluation of the pfaffian}

The pfaffian of a skew-symmetric matrix $R$ of dimension $2N$ and
with matrix elements $r_{ij}$ is defined as \cite{Caianello} \[
\textrm{pf}(R)=\frac{1}{2^{n}}\frac{1}{n!}\sum_{\textrm{Perm}}\epsilon(P)r_{i_{1}i_{2}}r_{i_{3}i_{4}}r_{i_{5}i_{6}}\ldots r_{2n-1,2n}\]
where the sum extends to all possible permutations of $i_{1},\ldots,i_{2n}$and
$\epsilon(P)$ is the parity of the permutation. For matrices of odd
dimension the pfaffian is by definition equal to zero. As an example,
the pfaffian of a $2\times2$ matrix $R$ is $\textrm{pf}(R)=r_{12}$
and for a $4\times4$ one $\textrm{pf}(R)=r_{12}r_{34}-r_{13}r_{24}+r_{14}r_{23}$.
Similarly to the case of determinants, exchanging rows $i$ and $j$
and the same time columns $i$ and $j$, multiplies the pfaffian by
minus one. Other useful properties of the pfaffian are

\begin{equation}
\textrm{pf}(P^{T}RP)=\textrm{det}(P)\textrm{pf}(R),\label{eq:PTRP}\end{equation}

\begin{equation}
\text{\textrm{pf}}\left(\begin{array}{cc}
0 & R\\
-R^{T} & 0\end{array}\right)=(-1)^{N(N-1)/2}\det(R)\label{eq:pfRC}\end{equation}
\begin{equation}
\textrm{pf }(R)=\left(\det R\right)^{1/2}\label{eq:pfRdetR}\end{equation}
A useful formula to compute pfaffians of small or simple matrices
is 
\begin{equation}
\textrm{pf}(R)=\sum_{j}(-1)^{i+j-1}r_{ij}\textrm{pf}(R_{ij})\label{eq:minor}\end{equation}
where $R_{ij}$ is the pfaffian-minor obtained by eliminating from
$R$ the two rows and two columns $i$ and $j$. 

The pfaffian of a complex skew-symmetric matrix $R$ of dimension
2N is evaluated numerically by first reducing the matrix to tridiagonal
form $R_{T}$. This reduction is accomplish by means of a set of 2(N-1)
successive Householder transformations $P_{i}$ exactly in the same
way as in the standard reduction of a symmetric matrix to tridiagonal
form \cite{Golub.96}. We have $P_{2(N-1)}\ldots P_{2}P_{1}RP_{1}^{T}P_{2}^{T}\ldots P_{2(N-1)}^{T}=R_{T}$
with \[
R_{T}=\left(\begin{array}{cccccc}
0 & r_{1} & 0 & 0 & \ldots & 0\\
-r_{1} & 0 & r_{2} & 0 & \ldots & 0\\
0 & -r_{2} & 0 & \ddots & \ldots & \vdots\\
0 & 0 & \ddots & 0 & \ddots & 0\\
\vdots & \vdots & \cdots & \ddots & 0 & r_{2N-1}\\
0 & 0 & \cdots & 0 & -r_{2N-1} & 0\end{array}\right)\]
where the special structure of a skew-symmetric and tridiagonal matrix
is evident. Using now Eq. (\ref{eq:PTRP}) we obtain $\det(P_{1})\ldots\det(P_{2(N-1)})\textrm{pf}(R)=\textrm{pf}(R_{T})$.
As Householder matrices are hermitian, unitary and have determinant
$\det(P_{i})=-1$ we finally arrive to $\textrm{pf}(R)=\textrm{pf}(R_{T})$.
To evaluate the pfaffian of the tridiagonal matrix we use the minor
expansion of Eq. (\ref{eq:minor}) that gives $\textrm{pf}(R_{T})=r_{1}r_{3}\ldots r_{2N-1}=\prod_{i=1}^{N}r_{2i-1}$.

\section{Derivation of the standard formula for the trace}

In this appendix the standard result of \cite{Rossignoli.94} for
the trace of a density operator is deduced from Eq. \eqref{eq:TrDFinal}.
I start considering \[
\tilde{\mathbb{M}}=\left(\begin{array}{cc}
T_{12}T_{22}^{-1} & -(T_{22}^{T}+\openone)\\
(T_{22}+\openone) & T_{21}T_{22}^{T}\end{array}\right)\]
Using Eq. \eqref{eq:pfRdetR} the pfaffian of $\mathbb{\tilde{M}}$
is written as the square root of its determinant, $\textrm{pf}\mathbb{\tilde{M}}=\left(\det\tilde{\mathbb{M}}\right)^{1/2}$.
By exchanging rows and columns conveniently\[
\det\tilde{\mathbb{M}}=(-)^{N}\left(\begin{array}{cc}
(T_{22}+\openone) & T_{21}T_{22}^{T}\\
T_{12}T_{22}^{-1} & -(T_{22}^{T}+\openone)\end{array}\right)\]
and applying the formula of a bipartite determinant $\det\left(\begin{array}{cc}
A & B\\
C & D\end{array}\right)=\det A\det(D-CA^{-1}B)$ the following result is obtained \begin{eqnarray*}
\det\tilde{\mathbb{M}} & = & \det T_{22}\det(\openone+T_{22})\\
 & \times & \det\left(\openone+\left(T_{22}^{T}\right)^{-1}+T_{12}T_{22}^{-1}(\openone+T_{22})^{-1}T_{21}\right).\end{eqnarray*}
But $T_{11}=\left(T_{22}^{T}\right)^{-1}+T_{12}T_{22}^{-1}T_{21}$
and $T_{22}^{-1}\left[(\openone+T_{22})^{-1}-\openone\right]=-(\openone+T_{22})^{-1}$
so
that\begin{eqnarray*}
\det\tilde{\mathbb{M}} & = & \det T_{22}\det(\openone+T_{22})\\
 & \times & \det\left(\openone+T_{11}-T_{12}(\openone+T_{22})^{-1}T_{21}\right)\\
 & = & \det T_{22}\det\left(\begin{array}{cc}
(T_{11}+\openone) & T_{12}\\
T_{21} & (T_{22}+\openone)\end{array}\right)\end{eqnarray*}
When the pfaffian is written as the square of the determinant the
sign is lost and therefore phases are irrelevant in the derivation.
Taking all this into account the result \[
\textrm{Tr}(\hat{\mathcal{D}})=\left[\det\left(\begin{array}{cc}
(T_{11}+\openone) & T_{12}\\
T_{21} & (T_{22}+\openone)\end{array}\right)\right]^{1/2}\]
is obtained up to a sign, which is the sought formula of Ref \cite{Rossignoli.94}.

\end{document}